\documentstyle[preprint,prb,aps]{revtex}
\begin{document}
\draft
\title{Partition asymptotics from 1D quantum entropy and energy 
currents}
\author{Miles P. Blencowe\cite{auth} and Nicholas C. Koshnick}
\address{Department of Physics and Astronomy, Dartmouth College, 
Hanover, New Hampshire 03755}
\date{\today}
\maketitle
\begin{abstract}
We give an alternative method to that of 
Hardy-Ramanujan-Rademacher to derive the leading exponential term 
in the asymptotic approximation to the partition function $p(n,a)$, 
defined as 
the number of decompositions of a positive integer $n$ into integer summands, 
with each 
summand appearing at most $a$ times in a given decomposition. 
The derivation involves mapping to an equivalent physical problem 
concerning the  quantum entropy and energy 
currents of particles flowing in a 
one-dimensional (1D) channel connecting thermal reservoirs, and which obey 
Gentile's intermediate 
statistics with 
statistical parameter $a$. The method is also applied to 
partitions associated with Haldane's fractional exclusion statistics.                 
\end{abstract}
\vfill
\eject

\section*{}
A classic result  in the theory of partitions is the 
Hardy-Ramanujan-Rademacher formula for the unrestricted 
partition function $p(n,\infty)$, wherein the latter, combinatoric 
quantity is represented as a power series whose terms involve 
elementary functions of $n$.\cite{hardy,rademacher,andrews} This 
series yields the following asymptotic approximation:
\begin{equation}
   p(n,\infty)\sim\frac{1}{4\sqrt{3} n} e^{\pi\sqrt{2/3}\sqrt{n}}.
    \label{asympt0} 
\end{equation}
A series representing $p(n,1)$, the number of decompositions of $n$ 
into distinct summands, has also been derived (see, e.g., Sec. 24.2.2 of 
Ref. \onlinecite{abramowitz}),
yielding the asymptotic approximation
\begin{equation}
   p(n,1)\sim\frac{1}{4\cdot 3^{1/4}\cdot n^{3/4}} 
   e^{\pi\sqrt{1/3}\ \sqrt{n}}.
    \label{asympt1} 
\end{equation}
And more recently,\cite{hagis} Hagis used the Hardy-Ramanujan-Rademacher 
method to derive a power series representation of $p(n,a)$ for 
arbitrary $a=1,2,\ldots$, yielding the asymptotic approximation
\begin{equation}
   p(n,a)\sim\frac{\sqrt{12}\ a^{1/4}}{(1+a)^{3/4}\ (24 n)^{3/4}} 
   e^{\pi\sqrt{2a/[3(1+a)]}\ \sqrt{n}},
    \label{asympt2} 
\end{equation}
where $n\gg a$. As an example, for $a=4$ the  
number of partitions of $n=1000$ to five significant figures 
is $2.4544 \times 10^{28}$,
while approximation (\ref{asympt2}) gives $2.4527 \times 10^{28}$, 
accurate to within $0.1\%$.

In the present work, we give an alternative and more direct derivation 
of the asymptotic approximation to $\ln p(n,a)$ which, from Eq. (\ref{asympt2}), 
is:
\begin{equation}
    \ln p(n,a) \sim \pi\sqrt{\frac{2 a}{3(1+a)}}\cdot\sqrt{n}.
    \label{lnasympt}
\end{equation}
The derivation 
begins by considering a 1D quantum channel which supports 
particles obeying  Gentile's intermediate statistics\cite{gentile} 
characterised by 
statistical parameter $a$, the maximum occupation number of 
particles in a single particle state, with $a=1$ describing fermions 
and $a=\infty$ bosons. The left end of the channel is connected to a 
particle source and 
the right end to a particle sink. The channel is dispersionless 
so that particle packets with different mean energies have the same 
velocity $c$ and hence  
transmission time ${\tau}=L/c$, where $L$ is the 
channel length. Imposing periodic boundary conditions on the channel 
length, the single-particle energies are $\epsilon_{j}=h f_{j} =h j/{\tau}$, 
$j=1,2,\dots$, where $h$ is Planck's constant. The
total energy $E_{n}$ of a given Fock state is 
$E_{n}=\sum_{j} \epsilon_{j} 
n_{j} ={nh}/{\tau}$, where $n=\sum_{j=1}^{\infty}j n_{j}$, and 
$n_{j}\leq a$ 
is the occupation number of, say, the right-propagating mode $j$. 

We now suppose that the  source emits a finite number of 
particles with fixed total energy $E_{n}$. The maximum possible 
entropy of this collection of right-propagating particles  
subject to the fixed energy constraint is $S(n,a)=k_{B}\ln 
p(n,a)$. Thus, the problem to determine the asymptotic approximation 
to $\ln p(n,a)$ is equivalent to determining the asymptotic approximation 
to the entropy $S(n,a)$ of the just-described physical system.
(C.f. Sec. 4 of Ref. \onlinecite{caves}, 
where the same set-up  restricted to bosons was considered in the 
problem to determine the optimum capacity for classical information transmission 
down a quantum channel.)

The crucial next step is to consider a slightly different set-up, in 
which the particle source and sink are replaced by two thermal  
reservoirs described by grand canonical ensembles, with the chemical 
potentials of the left and right reservoirs  
satisfying $\mu_{L}=\mu_{R}=0$, the 
temperature of the right 
reservoir  $T_{R}=0$,
and the temperature $T_{L}$ of the left reservoir chosen such that the 
thermal-averaged energy current flowing in the channel satisfies 
$\dot{\bar{E}}(T_{L},a)=E_{n}/{\tau}$. (Note 
that the chemical potentials are set to zero since there is no 
constraint on the thermal-averaged 
particle number.) 
With this choice, the thermal-averaged, channel entropy 
current $\dot{\bar{S}}(T_{L},a)$ coincides with $S(n,a)/{\tau}$ in the 
thermodynamic limit $E_{n}$ (equivalently 
$n)\rightarrow\infty$.

The advantage with using the latter, grand canonical ensemble description as 
opposed to the former,  microcanonical ensemble description is the greater 
ease with which the energy and entropy currents can be calculated.  
The starting formula for the
single channel energy current is:
\begin{equation}
    \dot{\bar{E}}(T,a)=\sum_{j=1}^{\infty}  \epsilon_{j} 
    \left[\bar{n}_{a}(\epsilon_{j})/L\right]c,
    \label{startenergy}
\end{equation}
where we have dropped the subscript on $T_{L}$,  and where 
$\bar{n}_{a}(\epsilon)$ 
is the intermediate statistics thermal-averaged occupation number of the 
right-moving state 
with energy $\epsilon$:\cite{gentile}
\begin{equation}
    \bar{n}_{a}(\epsilon)=\frac{1}{e^{\beta E}-1}-\frac{a+1}{e^{\beta E (a+1)}-1}.
    \label{ioccup}
\end{equation}
In the limit $L\rightarrow\infty$ (equivalently 
$\tau\rightarrow\infty$), we can replace the sum with an integral 
over $j$ and, 
changing integration variables $j\rightarrow \epsilon = (h/{\tau}) j=(hc/L) j$, we have  
[c.f. Eq. (13) of Ref. \onlinecite{sivan}]:
\begin{equation}
    \dot{\bar{E}}(T,a)=\frac{1}{h}
    \int_{0}^{\infty}d\epsilon \epsilon \bar{n}_{a}(\epsilon).
    \label{intenergy}
\end{equation}
A formula for entropy current can be derived as follows.
First note that the 
thermal-averaged occupation energy $\bar{\epsilon}=
\epsilon\bar{n}_{a}(\epsilon)$   and 
the entropy  $\bar{s}$ for a given mode with energy $\epsilon$ are 
related through the first law:  $d\bar{s}/dT 
=(1/T)d\bar{\epsilon}/dT$. Integrating with respect to temperature and 
then summing over the right propagating channel modes, we obtain    
\begin{equation}
    \dot{\bar{S}}(T,a)=-\frac{k_{B}}{h}
    \int_{0}^{\infty}d\epsilon \epsilon
    \int_{beta}^{\infty} d\beta' \beta'\frac{\partial \bar{n}_{a}}
    {\partial\beta'}.
    \label{intentropy}
\end{equation}

The integrals are straightforwardly carried out by  noting 
from (\ref{ioccup}) that the 
thermal-averaged occupation energy $\bar{\epsilon}=
\epsilon\bar{n}_{a}(\epsilon)$ of 
level $\epsilon$ for statistical parameter $a$ is just the difference in the 
thermal-averaged occupation energies of levels $\epsilon$ and $\epsilon(a+1)$ for 
bosons. Thus, we require only the integrals for the bosonic case: 
$\dot{\bar{E}}(T,\infty)=\pi^{2}(k_{B}T)^{2}/(6h)$ and 
$\dot{\bar{S}}(T,\infty)=\pi^{2}k_{B}^{2}T/(3h)$, giving
\begin{equation}
   \dot{\bar{E}}(T,a)=\left(1-\frac{1}{1+a}\right)\frac{\pi^{2}(k_{B}T)^{2}}{6h} 
   \label{intenergyf}
\end{equation}
and
\begin{equation}
   \dot{\bar{S}}(T,a)=\left(1-\frac{1}{1+a}\right)\frac{\pi^{2}k_{B}^{2} T}{3h}.
   \label{intentropyf}
\end{equation}

Comparing powers of $T$ appearing in Eqs. (\ref{intenergyf}) and 
(\ref{intentropyf}), and recalling that $\dot{\bar{E}}(T,a)=E_{n}/{\tau}$ 
and $\dot{\bar{S}}(T,a) \sim S(n,a)/{\tau}$,  we learn immediately that $\ln 
p(n,a) \sim C(a)\sqrt{n}$, where the $n$-independent factor $C(a)$ is 
given by
\begin{equation}
    C(a)=\frac{\sqrt{h}\dot{\bar{S}}(T,a)}{k_{B}\sqrt{\dot{\bar{E}}(T,a)}}.
    \label{theformula}
\end{equation}
Substituting in the expressions (\ref{intenergyf}) and
(\ref{intentropyf}) for $\dot{\bar{E}}$ and $\dot{\bar{S}}$, respectively, we 
finally obtain $C(a) = \pi\sqrt{2 a/[3(1+a)]}$, in agreement with 
Eq. (\ref{lnasympt}).

We will now carry out the same steps as above for particles 
obeying Haldane's fractional exclusion statistics\cite{haldane} to derive 
the asymptotic approximation to the logarithm of yet another type of 
partition function, $\tilde{p}(n,g)$, which also interpolates between the 
unrestricted 
and distinct partition functions [Eqs. (\ref{asympt0}) and (\ref{asympt1}), 
respectively]. Following the usual conventions, the statistics parameter is 
denoted by
$g=1/a$ (so that $g=0$ describes bosons and $g=1$ fermions).  
Partitions associated with exclusion statistics 
are subject to additional constraints as compared with partitions 
associated with intermediate 
statistics (see below). 

The energy and entropy currents for particles obeying exclusion 
statistics are\cite{rego,krive}
\begin{equation}
    \dot{\bar{E}}(T,g)=\frac{(k_{B}T)^{2}}{h}
    \int_{0}^{\infty}dx x \bar{n}_{g}(x)
    \label{halenergy}
\end{equation}
and
\begin{eqnarray}
    \dot{\bar{S}}(T,g)=&-&\frac{k_{B}^{2}T}{h}
    \int_{0}^{\infty}dx
    \left \{ \bar{n}_{g}\ln \bar{n}_{g}+(1-g \bar{n}_{g}) \ln(1-g \bar{n}_{g})\right.\cr
    &-&\left. [1+(1-g) \bar{n}_{g}] 
    \ln [1+(1-g) \bar{n}_{g}]\right \},
    \label{halentropy}
\end{eqnarray}
where $x=\beta \epsilon$ and the thermal-averaged occupation number is\cite{wu}
\begin{equation}
    \bar{n}_{g}(x)=
    \left[w(x)+g\right]^{-1},
    \label{hoccup1}
\end{equation}
with the function $w(x)$ given by the implicit equation
\begin{equation}
    w(x)^{g}[1+w(x)]^{1-g}=e^{x}.
    \label{hoccup2}
\end{equation}

Again, comparing powers of $T$ appearing in Eqs. (\ref{halenergy}) and 
(\ref{halentropy}), we learn immediately that $\ln 
\tilde{p}(n,g) \sim \tilde{C}(g)\sqrt{n}$, where the $n$-independent factor 
$\tilde{C}(g)$ is 
given in terms of $\dot{\bar{E}}$ and $\dot{\bar{S}}$ as in Eq. (\ref{theformula}). 
Substituting in the expressions for $\dot{\bar{E}}$ and $\dot{\bar{S}}$ and 
performing a change of variables from $x$ to $w$,\cite{rego} 
Eq. (\ref{theformula}) becomes after some algebra 
\begin{equation}
    \tilde{C}(g)=\frac{s(g)}{\sqrt{e(g)}},
    \label{theformula2}
\end{equation}
where
\begin{equation}
    e(g)= \int_{w_{g}(0)}^{\infty} dw 
    \frac{1}{w(1+w)}\left[(1-g)\ln (1+w) +g\ \ln w\right]
    \label{halenergyf}
\end{equation}
and
\begin{equation}
   s(g)= \int_{w_{g}(0)}^{\infty}dw 
    \left[\ln(1+w)/w-\ln w/(w+1)\right].
    \label{halentropyf}
\end{equation}
Using the identity $s(g)=2 
e(g)$, Eq. (\ref{theformula2}) can be further simplified to
\begin{equation}
    \tilde{C}(g)=\sqrt{2 s(g)}.
    \label{theformula3}
\end{equation}

Let us now
describe some of the properties and consequences of result 
(\ref{theformula3}). Integral 
(\ref{halentropyf}) can  be  rewritten in terms of dilogarithms 
\cite{lee} 
 and 
only for certain choices of  lower integration 
limit do  closed-form solutions  exist. For 
example, from (\ref{hoccup2}) we have $w_{g=0}(0)=0$ and 
$w_{g=1}(0)=1$ and solving the respective integrals, we obtain  
$s(0)=\pi^{2}/3$ and $s(1)=\pi^{2}/6$. Substituting 
these values into (\ref{theformula3}), we indeed obtain  the  
arguments 
of the exponentials in the asymptotic approximations to the unrestricted and 
distinct partition functions,  Eqs. (\ref{asympt0}) and (\ref{asympt1}) 
respectively. It is tempting to speculate that closed-form solutions 
to the integral $s(g)$  exist only for $g=0$, $1/2$, 
$1/3$, $1/4$, and $1$ in the interval $[0,1]$, 
since  it is only for these  rational
values that Eq. (\ref{hoccup2}) 
can be solved analytically for the lower integration limit 
$w_{g}(0)$.  For $g=1/2$, we have $w_{1/2}(0)=(-1+\sqrt{5})/2$ and 
$s(1/2)=\pi^{2}/5$, so that
\begin{equation}
    \ln \tilde{p}(n,1/2) \sim \pi\sqrt{2/5}\cdot\sqrt{n}.
    \label{lnasympt2}
\end{equation}
Note that $\tilde{C}_{g=1/2}\ (=\pi\sqrt{2/5})< C_{a=2}\ (=2\pi/3)$, 
signalling the fact that $\tilde{p}(n,g)<p(n,a=1/g)$ for $0<g<1$, a 
consequence of additional constraints  on the allowed 
partitions associated with Haldane's statistics. These constraints are discussed 
in Ref.\ \onlinecite{murthy}.      
The above, closed-form solutions for $g=0$, $g=1/2$, and $1$
were obtained by solving the integral $s(g)$ 
numerically and then noting that the result when divided by $\pi^{2}$ 
was rational.  This method does not work for
the $g=1/3,1/4$ cases, however, owing to the complicated form of the lower 
limits $w_{1/3}(0)$ and $w_{1/4}(0)$ (they are roots of third and fourth 
degree polynomial equations, respectively). A more sophisticated 
method is required in order to determine whether or not closed-form 
solutions exist for these latter two cases.

\acknowledgements
We would like to thank Peter Hagis Jr. and George Andrews for very 
helpful correspondences.
Discussions with Makoto Itoh and Jay Lawrence are also gratefully 
acknowledged.

\end{document}